\documentclass[prl,aps,superscriptaddress,twocolumn]{revtex4-1}
\setcitestyle{super}
\usepackage{amsfonts,amssymb,amscd,amsthm}
\usepackage{graphicx}
\usepackage{threeparttable}
\usepackage{multirow}
\usepackage{mathrsfs}
\usepackage[intlimits]{amsmath}
\usepackage{array}
\usepackage{xcolor}
\usepackage[colorlinks, citecolor=red]{hyperref}
\usepackage{lineno}
\usepackage{etoolbox}
\usepackage{color,soul}
\usepackage[normalem]{ulem}

\definecolor{blue}{rgb}{0,0,1}
\definecolor{red}{rgb}{1,0,0}
\definecolor{green}{rgb}{0,1,0}

\newcommand{\ket}[1]{\ensuremath{\left|#1\right\rangle}}

\begin{document}

\title{Breaking the scalability barrier via a vertical tunable coupler in 3D integrated transmon system}

\author{Xudong Liao}
\altaffiliation{These three authors contributed equally to this work.}
\affiliation{National Laboratory of Solid State Microstructures, School of Physics, Nanjing University, Nanjing, 210093 Jiangsu, China}
\affiliation{Shishan Laboratory, Nanjing University, Suzhou, 215163 Jiangsu, China}
\affiliation{Jiangsu Key Laboratory of Ouantum Information Science and Technology, Nanjing University, Suzhou, 215163 Jiangsu, China}

\author{Shuyi Pan}
\altaffiliation{These three authors contributed equally to this work.}
\affiliation{National Laboratory of Solid State Microstructures, School of Physics, Nanjing University, Nanjing, 210093 Jiangsu, China}
\affiliation{Shishan Laboratory, Nanjing University, Suzhou, 215163 Jiangsu, China}
\affiliation{Jiangsu Key Laboratory of Ouantum Information Science and Technology, Nanjing University, Suzhou, 215163 Jiangsu, China}

\author{Zhenxing Zhang}
\altaffiliation{These three authors contributed equally to this work.}
\affiliation{Tencent Quantum Laboratory, Tencent, Shenzhen, Guangdong 518057, China}

\author{Sainan Huai}
\altaffiliation{These three authors contributed equally to this work.}
\affiliation{Tencent Quantum Laboratory, Tencent, Shenzhen, Guangdong 518057, China}

\author{Zhiwen Zong}
\affiliation{Tencent Quantum Laboratory, Tencent, Shenzhen, Guangdong 518057, China}

\author{Xiaopei Yang}
\affiliation{Tencent Quantum Laboratory, Tencent, Shenzhen, Guangdong 518057, China}

\author{Kunliang Bu}
\affiliation{Tencent Quantum Laboratory, Tencent, Shenzhen, Guangdong 518057, China}

\author{Wen Zheng}
\affiliation{National Laboratory of Solid State Microstructures, School of Physics, Nanjing University, Nanjing, 210093 Jiangsu, China}
\affiliation{Shishan Laboratory, Nanjing University, Suzhou, 215163 Jiangsu, China}
\affiliation{Jiangsu Key Laboratory of Ouantum Information Science and Technology, Nanjing University, Suzhou, 215163 Jiangsu, China}

\author{Xinsheng Tan}
\affiliation{National Laboratory of Solid State Microstructures, School of Physics, Nanjing University, Nanjing, 210093 Jiangsu, China}
\affiliation{Shishan Laboratory, Nanjing University, Suzhou, 215163 Jiangsu, China}
\affiliation{Jiangsu Key Laboratory of Ouantum Information Science and Technology, Nanjing University, Suzhou, 215163 Jiangsu, China}

\author{Yang Yu}
\affiliation{National Laboratory of Solid State Microstructures, School of Physics, Nanjing University, Nanjing, 210093 Jiangsu, China}
\affiliation{Shishan Laboratory, Nanjing University, Suzhou, 215163 Jiangsu, China}
\affiliation{Jiangsu Key Laboratory of Ouantum Information Science and Technology, Nanjing University, Suzhou, 215163 Jiangsu, China}

\author{Yuan Li}
\email{stevenyli@tencent.com}
\affiliation{Tencent Quantum Laboratory, Tencent, Shenzhen, Guangdong 518057, China}

\author{Yi-Cong Zheng}
\email{yicongzheng@tencent.com}
\affiliation{Tencent Quantum Laboratory, Tencent, Shenzhen, Guangdong 518057, China}

\author{Tianqi Cai}
\email{tianqicai@tencent.com}
\affiliation{Tencent Quantum Laboratory, Tencent, Shenzhen, Guangdong 518057, China}

\author{Shengyu Zhang}
\email{shengyzhang@tencent.com}
\affiliation{Tencent Quantum Laboratory, Tencent, Shenzhen, Guangdong 518057, China}

\begin{abstract}
Scaling superconducting quantum processors beyond the constraints of monolithic planar architectures is essential for fault-tolerant quantum computation. Here we demonstrate a three-dimensional (3D) integrated superconducting quantum processor in which two qubit chips are vertically stacked on opposing sides of a carrier chip and galvanically connected via multilayer flip-chip bonding. Intrachip qubit coupling is mediated by planar tunable couplers, whereas interchip coupling is enabled by vertical tunable couplers embedded in the carrier chip. Randomized benchmarking reveals simultaneous single-qubit gate fidelities of 99.87\% with negligible crosstalk, and controlled-Z gates achieve an average fidelity of 97.5\% for both intrachip and interchip operations. We further demonstrate high-fidelity Bell-state preparation and coherent generation of a four-qubit $W$ state, confirming the architecture's capability for interchip entanglement distribution. These results establish vertical coupling as a promising pathway toward scalable quantum processors compatible with advanced quantum error-correcting codes.
\end{abstract}

\maketitle

\section{Introduction}\label{Sec1}

Fault-tolerant quantum computation demands millions of physical qubits, far beyond what monolithic single-chip scaling can deliver owing to rapidly declining fabrication yields, increasingly complex signal routing, and growing crosstalk with increasing chip area. Distributed modular quantum computing\cite{bravyi2022future, awschalom2021development} offers a promising alternative, and recent experiments have linked separate superconducting processors via coaxial cables for remote state transfer and entanglement generation\cite{roch2014observation, narla2016robust, dickel2018chip, campagne2018deterministic, kurpiers2018deterministic, axline2018demand, leung2019deterministic, PhysRevLett.125.260502, zhong2021deterministic, burkhart2021error, niu2023low, zhou2023realizing, mollenhauer2407high, qiu2025deterministic, qiu2025thermal, almanakly2025deterministic, song2025realization, li2025fast}. However, this approach suffers from practical constraints such as unavoidable interface losses and considerable space requirements within dilution refrigerators, motivating a shift toward more integrated modular processor designs\cite{brecht2016multilayer, rosenberg2020solid}.

Currently, qubit and control layers are typically distributed on separate chips and interconnected via flip-chip bonding\cite{rosenberg20173d, foxen2018qubit, niedzielski2019silicon, kosen2022building} to form an integrated processor. This approach allows multiple qubit chips to be simultaneously bonded to a single carrier chip with corresponding control lines\cite{gold2021entanglement, wu2024modular, norris2026performance, dalton2025resource}, and partitioning a large qubit array into smaller chips improves the fabrication yield of individual dies. However, as the number of qubits and chips grows, the densely packed control lines and qubit couplings on the carrier chip lead to increasing crosstalk, posing a significant challenge for scaling. One solution is to combine through-silicon vias (TSVs)\cite{gambino2015overview, yost2020solid} with flip-chip techniques to alleviate crosstalk and routing congestion by distributing signal paths across multiple layers\cite{brecht2016multilayer, rosenberg20173d}, but TSV fabrication which involves deep silicon etching and conformal metallization of high-aspect-ratio holes remains substantially more challenging than mature flip-chip processes. To overcome this challenge, an alternative route distributes different functional layers on both sides of a single chip\cite{rahamim2017double, bakr2025intrinsic}, significantly enhancing integration density. Despite these advances, the simultaneous suppression of crosstalk and scalability of qubits remains an open problem, and existing approaches are confined to lateral scaling parallel to the chip plane and have yet to exploit vertical integration perpendicular to the two-dimensional chip surface.

In this work, we propose and demonstrate a three-dimensional (3D) integrated modular superconducting quantum processor that enables tunable coupling between physically distinct modules in the vertical direction. In this device, two independent qubit chips are placed on opposing sides of a carrier chip and galvanically connected through a multilayer flip-chip bonding process. Within each qubit chip, nearest-neighbor qubits are coupled via conventional planar tunable couplers\cite{yan2018tunable}, while qubits on different chips are coupled through vertical tunable couplers on the carrier chip, enabling independently controlled interchip entangling operations. Experimental characterization demonstrates that the multilayer 3D integration process is fully compatible with high-performance superconducting qubits. To benchmark vertical couplers against planar couplers, we performed both isolated and simultaneous randomized benchmarking (RB)\cite{knill2008randomized} on four representative qubits. The simultaneous single-qubit gate fidelity averages 99.87\%, in excellent agreement with the isolated fidelity of 99.88\%, confirming that crosstalk mediated by both types of couplers is well suppressed at the operating point. For two-qubit operations, both intrachip and interchip controlled-Z (CZ) gates achieve an average fidelity of 97.5\%, validating the feasibility of high-fidelity two-qubit gate operations through the vertical coupling scheme. To further characterize entanglement generation, we prepared Bell states with a fidelity of 97.0\% using calibrated CZ gates, with entanglement quality confirmed by quantum state tomography (QST)\cite{roos2004bell, steffen2006measurement}. A joint-evolution experiment with four qubits demonstrates the multiqubit cooperative capability of this architecture, achieving preparation of a four-qubit $W$ state within approximately 68 ns and providing compelling evidence for coherent interchip control of multiqubit entanglement. Furthermore, this architecture offers a natural pathway toward larger-scale quantum processors through extended interchip connections and additional vertical coupler layers. It also holds promise for implementing complex quantum error-correcting codes, such as three-dimensional stacked surface codes\cite{min20263d} and quantum low-density parity-check (qLDPC) codes\cite{mathews2025placing}, thereby laying a foundation for fault-tolerant quantum computation.

\section{Device design and fabrication}\label{Sec2}

\begin{figure}[tb]
\includegraphics{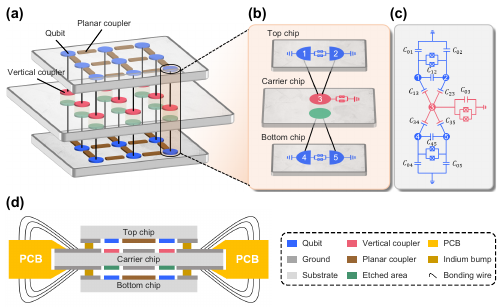}
\caption{\textbf{Schematics illustrating the 3D integrated transmon system.} (a) Overall schematic of the system. Qubits (blue) reside on the top and bottom chips, with nearest neighbors coupled via planar couplers (brown). Vertical couplers (red) and the corresponding etched regions (green) are fabricated on opposing sides of the carrier chip to couple qubits on separate chips, as indicated by the black solid lines. (b) Schematic of the vertical coupling unit cell highlighted by the orange cylindrical region in (a). (c) Equivalent circuit of the structure shown in (b). Circuit node labels correspond to the pad labels in (b). (d) Simplified cross-sectional illustration of the 3D integrated quantum processor. Light gray regions represent chip substrates, while dark gray regions represent the ground planes. The carrier chip is galvanically connected to the top and bottom chips via indium bumps (dark gold), and the entire integrated chip is wire-bonded (black arcs) to the PCB (bright gold) for packaging.}
\label{fig:Fig1}
\end{figure}

The architecture of the 3D integrated transmon system is illustrated in Fig. \ref{fig:Fig1}(a). The top and bottom chips, each containing the qubit layer, are placed on opposing sides of the carrier chip to form a sandwich-type stacked structure. Within each qubit chip, nearest-neighbor qubits are coupled via planar tunable couplers, while vertical couplers on the carrier chip mediate interchip qubit-to-qubit interactions with corresponding regions on the opposite side etched away to enhance capacitive coupling. The overall coupling architecture in both the lateral and vertical directions follows the qubit-coupler-qubit (QCQ) scheme\cite{yan2018tunable, sete2021floating}, ensuring that every qubit-to-qubit interaction is mediated by an independently tunable coupler. 

A representative unit cell along the vertical direction is shown in Fig. \ref{fig:Fig1}(b). Qubits on the top chip couple capacitively and directly to the vertical couplers, while qubits on the bottom chip couple to the vertical couplers through the etched regions. The equivalent circuit is depicted in Fig. \ref{fig:Fig1}(c). Both planar and vertical couplers are designed to activate two-qubit gates and to fully suppress residual qubit-to-qubit couplings when idling. Consequently, any qubit can interact with its neighbors in both the lateral and vertical directions, or even simultaneously in both, greatly enhancing the integration density of the processor. Since control lines are routed on both sides of the carrier chip, the packaging PCB incorporates a hollow structure to accommodate the entire integrated chip, as shown in Fig. \ref{fig:Fig1}(d). The top and bottom chips are galvanically connected to the carrier chip via indium bumps\cite{foxen2018qubit}, and the integrated assembly is wire-bonded to the PCB to interface the double-sided control lines with external electronics.

\begin{figure*}[htb]
\includegraphics{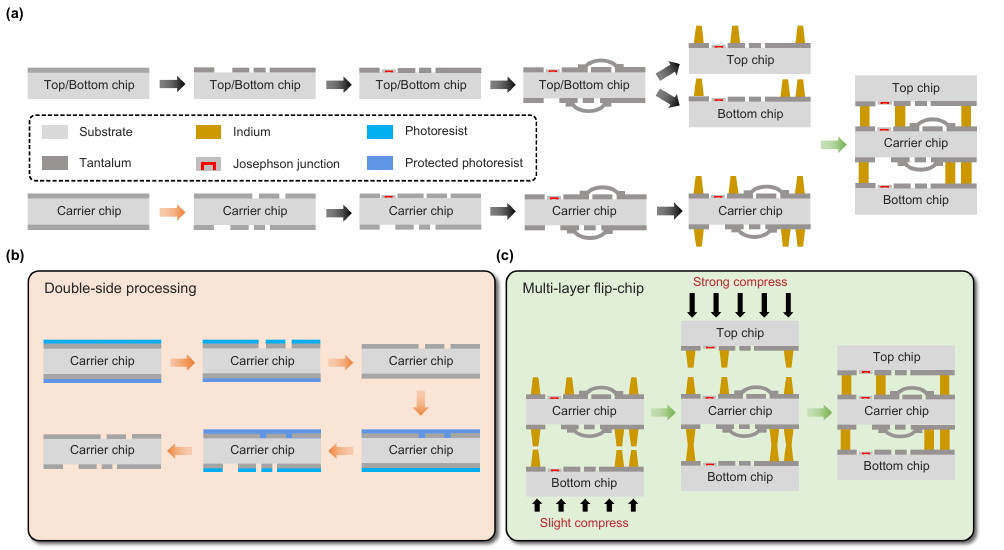}
\caption{\textbf{Chip fabrication procedures.} (a) Overview of the key fabrication steps. Base CPW layers are defined by optical lithography and etching, followed by Josephson-junction formation via EBL and double-angle evaporation. Airbridge fabrication for the top and bottom chips is optional (not included in our device). Indium bumps are then evaporated, and the final integrated chip is assembled by multilayer flip-chip bonding. (b) Double-sided processing scheme [orange arrow in (a)]. Optical lithography and etching are carried out on one side of the carrier chip while the opposite side is protected with photoresist, and the procedure is then repeated for the other side. (c) Multilayer flip-chip bonding scheme [green arrow in (a)]. The bottom chip is first bonded to the carrier chip with a slight compression, and the resulting assembly is then bonded to the top chip with a strong compression to yield the final integrated chip.}
\label{fig:Fig2}
\end{figure*}

All three chips share the same base material, 200 nm-thick $\alpha$-phase tantalum on sapphire substrate. The top and bottom chips were fabricated using standard superconducting qubit processes\cite{chen2018metrology, foxen2018qubit}, comprising coplanar waveguide (CPW) layer patterning, Josephson-junction formation by electron-beam lithography (EBL) and double-angle evaporation\cite{kreikebaum2020improving}, and deposition of 8 $\mu$m-thick indium bumps. In addition, indium spacers were deposited at the four corners of both the top and bottom chips to control the chip-to-chip gap during subsequent flip-chip bonding. Fabrication of the carrier chip is more involved because structures must be defined on both sides of the substrate. To enhance the capacitive coupling between qubits on opposing chips, a thinner 330 $\mu$m-thick sapphire substrate was chosen for the carrier chip. Because both sides of this substrate carry functional structures, each side was patterned sequentially [Fig. \ref{fig:Fig2}(b)]: photoresist was spin-coated on the side to be patterned, while a protective photoresist layer was applied to the opposite side to prevent damage during processing. Optical lithography and reactive-ion etching (RIE) were then performed on the patterning side. After stripping both photoresist layers to reveal the CPW structures, the same procedure was repeated for the opposite side, yielding double-sided CPW layers. Similarly, since Josephson junctions are required only for the vertical couplers on the top side, we applied protective photoresist to the bottom side and then performed EBL and double-angle evaporation exclusively on the top side. Tantalum airbridges on both sides were realized by combining the double-sided processing technique with standard airbridge fabrication procedures\cite{bu2025tantalum}. Finally, 8 $\mu$m-thick indium bumps were evaporated onto both sides of the carrier chip.

To assemble the three chips into a single galvanically connected unit, we developed a multilayer flip-chip bonding process [Fig. \ref{fig:Fig2}(c)]. The bottom chip was first aligned and bonded to the bottom side of the carrier chip using a slight compression. This gentle initial bond deforms the indium bumps just enough to establish reliable galvanic contact while preserving sufficient bump height for the subsequent bonding step. The resulting two-chip assembly was then flipped and bonded to the top chip with a stronger compression to ensure robust galvanic connections at both interfaces. Because the indium bumps on both the top and bottom chips share identical initial height, this two-step protocol with carefully calibrated compression forces yields uniform interchip gaps of approximately 8 $\mu$m at both the top and bottom interfaces.

\section{Experimental results}\label{Sec3}

\begin{figure*}[htb]
\includegraphics{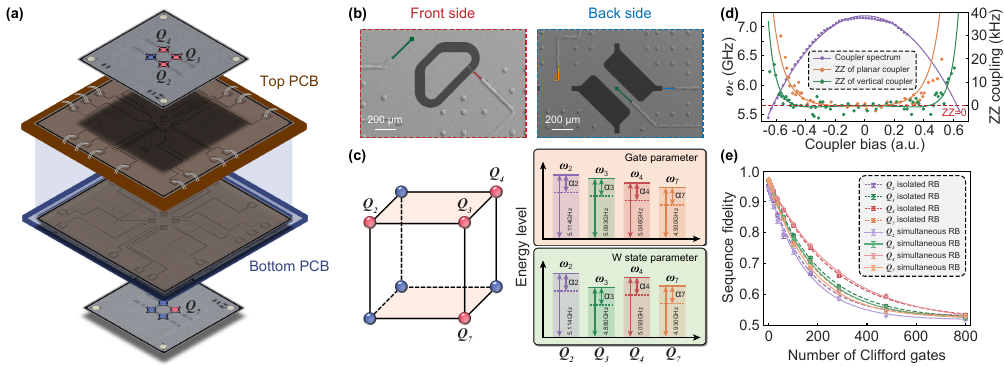}
\caption{\textbf{Basic device characterization.} (a) Optical micrograph of the device. The uppermost and lowermost gray planes correspond to the top and bottom chips, each containing qubits and readout resonators. The two central taupe-colored planes represent the two sides of the carrier chip, comprising vertical couplers, etched regions, and control lines. (b) Magnified SEM images of both sides of the carrier chip. The left panel (red dashed box) shows the front side housing the vertical couplers, and the right panel (blue dashed box) shows the back side with the corresponding etched regions. Control lines are color-coded: blue and green for qubit $XY$ and $Z$ control lines, respectively; red for $Z$ line of the vertical coupler; and yellow for $Z$ line of the planar coupler. (c) Left panel: topological connectivity diagram of qubits on the integrated chip. Orange planes represent the upper and lower qubit layers depicted in (a). Solid and dashed black lines indicate couplings between qubits, with red (blue) spheres denoting qubits used (not used) in our experiment. Right panel: the orange panel shows the biased qubit frequencies used for single-qubit and two-qubit gate calibrations, and the green panel shows those used for $W$-state preparation. (d) Energy spectrum (blue) of the vertical coupler between $Q_3$ and $Q_7$, and the residual $ZZ$ coupling strength for each qubit pair as a function of flux bias (orange for $Q_2-Q_3$ or $Q_3-Q_4$, green for $Q_3-Q_7$). (e) Isolated and simultaneous RB results for $Q_2$, $Q_3$, $Q_4$, and $Q_7$.}
\label{fig:Fig3}
\end{figure*}

The device layout is shown in Fig. \ref{fig:Fig3}(a). The uppermost and lowermost gray planes correspond to the top chip and bottom chips, respectively, each containing four qubits and their associated readout resonators. The two taupe-colored planes represent the two sides of the carrier chip, which contain vertical couplers, etched regions and control lines, as the magnified SEM images shown in Fig. \ref{fig:Fig3}(b).

To analyze this system in detail, we extract the qubit connectivity to form the topological diagram on the left of Fig. \ref{fig:Fig3}(c). The upper and lower orange planes represent the top and bottom chips, respectively. The spheres denote qubits, and black lines represent the tunable coupling between qubits. The four red spheres, labelled $Q_2$, $Q_3$, $Q_4$, and $Q_7$, represent the four qubits used in our experiment. The connections among these four qubits include both intrachip and interchip couplings. Therefore, the four qubits can be used to simultaneously calibrate and compare the modulation performance of both planar and vertical couplers. As shown in Fig. \ref{fig:Fig3}(d), we first measured the energy spectrum of the vertical coupler between $Q_3$ and $Q_7$, and subsequently characterized the residual ZZ coupling strengths among all four qubit pairs as a function of flux bias using Ramsey sequences. The results show that there exist ZZ closing points for every coupler.

\begin{figure}[htb]
\includegraphics{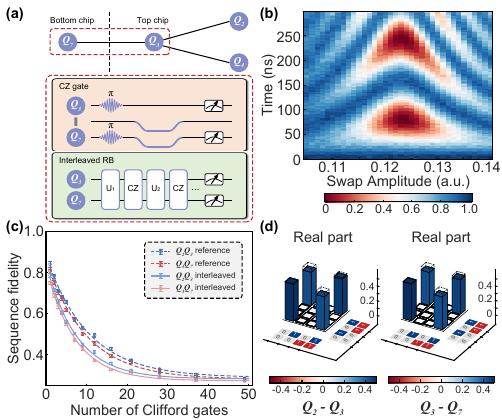}
\caption{\textbf{Two-qubit CZ gate experiments.} (a) Top: topological structure diagram of the qubits used in our experiments. Bottom: pulse sequences for CZ gate and interleaved RB. (b) Swap chevron patterns for the two-qubit $\ket{11}$–$\ket{02}$ resonance, which is used to calibrate diabatic CZ gate. (c) Interleaved RB of CZ gate for $Q_2$–$Q_3$ and $Q_3$–$Q_7$ pairs, achieving an average gate fidelity of 97.5\%. (d) Quantum state tomography of the Bell triplet state in $Q_2$–$Q_3$ and $Q_3$–$Q_7$ pairs, with an average state fidelity of 97\%.}
\label{fig:Fig4}
\end{figure}

To further assess the ZZ crosstalk between qubits, we biased the qubit frequencies to the idle points indicated by the orange panel in Fig. \ref{fig:Fig3}(c). Subsequently, we benchmarked the single-qubit gate fidelity using both isolated and simultaneous RB\cite{knill2008randomized}. With all couplers biased to their respective ZZ closing points [Fig. \ref{fig:Fig3}(d)], we find that the simultaneous single-qubit gate fidelity (99.87\%) exhibits only a marginal decrease relative to the isolated value (99.88\%) as shown in Fig. \ref{fig:Fig3}(e). This result indicates that both the planar and vertical couplers can effectively suppress ZZ crosstalk between qubits at the chosen operating points.

We next compare vertical- and planar-coupler-mediated CZ gates using the interchip $Q_3$--$Q_7$ and intrachip $Q_2$--$Q_3$ pairs, respectively. The corresponding pulse sequences are shown in Fig. \ref{fig:Fig4}(a). Taking the $Q_3$-$Q_7$ pair as an example, we first apply two $\pi$ pulses to prepare the state $\ket{11}$. The frequency of $Q_7$ is then tuned such that $\ket{11}$ is brought into resonance with $\ket{02}$, while the coupler is simultaneously modulated to activate the interaction. By sweeping the flux bias of $Q_7$, we obtained the swap-chevron pattern between $\ket{11}$ and $\ket{02}$ (or $\ket{20}$), as shown in Fig. \ref{fig:Fig4}(b). Interleaved RB is then performed for both the $Q_3$--$Q_7$ and $Q_2$--$Q_3$ CZ gates. The two gates show comparable performance, with an average CZ-gate fidelity of $97.5\%$, as shown in Fig. \ref{fig:Fig4}(c). The gate fidelity is mainly limited by the relatively short qubit relaxation times (the average $T_1$ is approximately $20~\mu\mathrm{s}$), and the comparatively long CZ-gate duration [as indicated by the chevron pattern in Fig. \ref{fig:Fig4}(b)] (the devide parameters and further optimization of gate operation can be found in supplementary materials). Finally, we prepared the Bell state $(\ket{00}+\ket{11})/\sqrt{2}$ for both the $Q_2$-$Q_3$ and $Q_3$-$Q_7$ pairs in Fig. \ref{fig:Fig4}(d) using the corresponding CZ pulse sequences, achieving an average QST (quantum state tomography) fidelity of 97\%. 

\section{Discussions}\label{Sec4}

To further demonstrate the multiqubit cooperative capability of the vertical coupling architecture, we extend the system beyond two-qubit entangling operations to explore both multipartite entanglement generation and the scalable potential of this 3D integration scheme. In the following, we present a four-qubit $W$-state generation experiment validating interchip multiqubit entanglement and discuss scaling strategies toward larger-scale processors.

\begin{figure}[htb]
\includegraphics{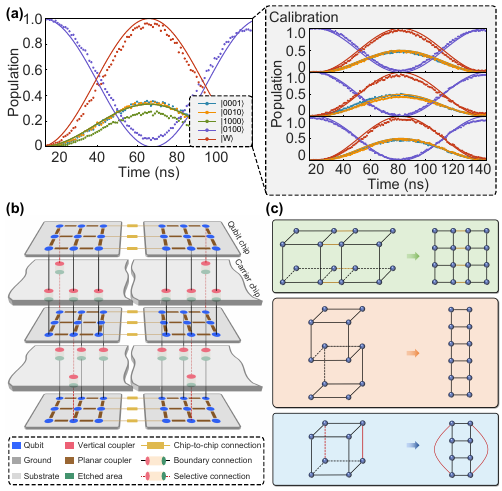}
\caption{\textbf{Joint-evolution experiment and scalable architecture.} (a) Left panel: dynamic evolution of four-qubit state $\ket{Q_2Q_3Q_4Q_7}$, realizing a four-qubit $W$ state at approximately 68 ns. Right panel: dynamic evolution of the three-qubit states $\ket{Q_2Q_3Q_7}$, $\ket{Q_3Q_4Q_7}$, and $\ket{Q_2Q_3Q_4}$, which is used to calibrate the four-qubit $W$-state experiment. (b) Envisioned scheme of scaling 3D integrated quantum processor. Quantum chips are connected laterally through chip-to-chip connections and vertically through vertical couplers on multiple stacked carrier chips. Vertical black solid lines indicate boundary connections while red dashed lines indicate selective connections. (c) Topological structure diagrams of three representative configurations in (b). Left: three-dimensional diagrams. Right: unfolded planar diagrams. Green and orange panel represent chip scalability in lateral and vertical direction, respectively, and blue panel represents selective connection scalability between chips.}
\label{fig:Fig5}
\end{figure}

We first perform a four-qubit $W$-state generation experiment as shown in Fig. \ref{fig:Fig5}(a). We measure the coherent oscillation between $\ket{100}_{Q_3Q_iQ_j}\ (i,j \ \text{in} \ [3,4,7])$ and $\ket{0}_{Q_3}\otimes(\ket{10}_{Q_iQ_j}+\ket{01}_{Q_iQ_j})/\sqrt{2}$, and use it to match the two coupling strengths, determining the working points where $g_{32} = g_{34} = g_{37}$. After this calibration, the system is initialized in $\ket{Q_2Q_3Q_4Q_7}=\ket{0100}$. The dynamics is effectively confined to a two-dimensional subspace spanned by $\ket{0100}$ and the three-qubit $W$ state (\(\ket{W_{247}}=(\ket{1000}+\ket{0010}+\ket{0001})/{\sqrt{3}}\)), leading to coherent oscillation between these two states. The pairwise equal-coupling calibration and the subsequent generation of $\ket{W_{247}}$ demonstrate the fine tunability and reliable control of the vertical coupler.

Furthermore, this connectivity naturally suggests a scalable route toward larger quantum processors. As illustrated in Fig. \ref{fig:Fig5}(b), the architecture can be expanded laterally by connecting multiple qubit chips through chip-to-chip connections\cite{gold2021entanglement, wu2024modular, norris2026performance, zhao2022tunable}, and vertically by stacking additional carrier-chip layers, where each layer can comprise multiple independent chips to improve fabrication yield. Moreover, by selectively activating the vertical couplers on different carrier layers, qubits residing on different chips can be connected on demand, thereby enabling a broad class of coupling topologies tailored to specific applications. Figure \ref{fig:Fig5}(c) illustrates three representative configurations, with the left column showing three-dimensional lattice structures and the right column displaying the corresponding unfolded planar diagrams. The green and orange panels highlight chip scalability in the lateral and vertical directions, respectively, while the blue panel represents selective connection scalability between different qubit chips. These examples demonstrate that the architecture supports flexible and extensible coupling topologies suitable for implementing complex quantum error-correcting codes, such as three-dimensional stacked surface codes\cite{min20263d} and qLDPC codes\cite{mathews2025placing}.

In summary, we have demonstrated a stacked 3D integrated superconducting quantum processor in which tunable vertical couplers provide an additional coupling degree of freedom. ZZ closing points are identified for all relevant qubit pairs, and both isolated and simultaneous RB confirm that ZZ crosstalk is well suppressed at the operating point. Both the intrachip $Q_2–Q_3$ and interchip $Q_3–Q_7$ CZ gates achieve an average fidelity of 97.5\%, with Bell-state tomography yielding an average state fidelity of 97\%. Notably, the near-identical performance of intrachip and interchip gates indicates that vertical connectivity introduces no appreciable degradation relative to conventional planar coupling. Looking forward, several challenges remain before this strategy can be extended to substantially larger systems: at the qubit level, the interface energy participation ratios\cite{wang2015surface, dial2016bulk, gambetta2016investigating, martinis2022surface, kosen2022building, smirnov2024wiring} should be further analyzed to minimize dielectric and interface losses; at the device level, residual ZZ interactions, flux-noise sensitivity, and coherence loss associated with the vertical couplers need to be reduced; and at the system level, denser control wiring, scalable packaging, efficient cross-layer thermalization, and automated calibration protocols will become increasingly important as the stack size grows. Nevertheless, the additional topological degree of freedom offered by vertical connectivity provides a promising hardware pathway toward fault-tolerant quantum computation.

\section{Acknowledgements} 
We thank Fuming Liu, Guanglei Xi, Qiaonian Yu and Hualiang Zhang for supporting room-temperature electronics. 

\section{Author Contributions}
T.Q.C. developed the idea and conceived the experiment. X.D.L., T.Q.C., and S.N.H. designed the device. Y.L. developed the fabrication method and fabricated the devices with assistance of T.Q.C., Z.X.Z., S.N.H., K.L.B. and X.P.Y.. T.Q.C., Z.X.Z., S.Y.P. and Z.W.Z. established the measurement setup and performed experimental measurements with assistance of S.N.H., X.D.L., X.P.Y., Y.C.Z. and S.Y.Z.. S.Y.P. and X.D.L. performed the theoretical analyses and numerical simulations with assistance of T.Q.C.. S.Y.P. and X.D.L wrote the manuscript with feedback from all authors. Y.L., T.Q.C., Y.C.Z. and S.Y.Z. supervised the project. All authors contributed to the discussion of the results and development of the manuscript.


%

\bigskip

\end{document}